# Unified understanding of superconductivity and Mott transition in alkali-doped fullerides from first principles

Yusuke Nomura[1*], Shiro Sakai[2], Massimo Capone[3], and Ryotaro Arita[2,4]

[1]Department of Applied Physics, University of Tokyo, Hongo, Bunkyo-ku, Tokyo, 113-8656, Japan

[2]Center for Emergent Matter Science (CEMS), RIKEN, 2-1, Hirosawa, Wako, Saitama 351-0198, Japan

[3]International School for Advanced Studies (SISSA) and CNR-IOM Democritos National Simulation Center, Via Bonomea 265, I-34136, Trieste, Italy

[4]JST ERATO Isobe Degenerate π-Integration Project, AIMR, Tohoku University, 2-1-1 Katahira, Aoba-ku, Sendai, 980-8577, Japan

[*]e-mail: yusuke.nomura@riken.jp

**One Sentence Summary:** A quantitative theory unveils unique synergy of phonons and electron correlations behind exotic pairing next to Mott phase.

**Abstract:** Alkali-doped fullerides $A_3C_{60}$ ($A$=K, Rb, Cs) are surprising materials where conventional phonon-mediated superconductivity and unconventional Mott physics meet, leading to a remarkable phase diagram as a function of the volume per $C_{60}$ molecule. Here we address these materials with a state-of-the-art calculation, where we construct a realistic low-energy model from first principles without using a priori information other than the crystal structure and solve it with an accurate many-body theory. Remarkably our scheme comprehensively reproduces the experimental phase diagram including the low-spin Mott-insulating phase next to the superconducting phase. Most remarkably, the critical temperature $T_c$'s calculated from first principles quantitatively reproduce the experimental values. The driving force behind the surprising phase diagram of $A_3C_{60}$ is a subtle competition between Hund's coupling and Jahn-Teller phonons, which leads to an effectively inverted Hund's coupling. Our results establish that the fullerides are the first members of a novel class of molecular superconductors in which the multiorbital electronic correlations and phonons cooperate to reach high-$T_c$ $s$-wave superconductivity.

# Introduction

The alkali-doped fullerides are unique exotic *s*-wave superconductors with maximum critical temperature $T_c \sim 40$ K, which is extremely high for the modest width of around 0.5 eV of the bands relevant to the superconductivity (1-9). Although the evidence for a phononic pairing mechanism is robust and may suggest a conventional explanation for superconductivity in these materials (2), the experimental observation of a Mott insulating phase next to the superconducting phase in $Cs_3C_{60}$ changes the perspective since the same strong correlations responsible for the Mott state are likely to influence also the superconducting state (3-9). This raises a fundamental question why *s*-wave superconductivity, which is induced by some attractive interaction, is so resilient to (or even enhanced by) the strong repulsive interactions (2). The fingerprints of phonons observed in the Mott insulating phase, such as the low-spin state (4-9) and the dynamical Jahn-Teller effect (10), confirm a non-trivial cooperation between the electron correlations and the electron-phonon interactions.

A variety of scenarios have been proposed to explain the superconductivity (2,11-21). However, even the most successful theories include some adjustable parameter and/or neglect, e.g., the realistic band structure, the intermolecular Coulomb interactions, and Hund's coupling. These assumptions have prevented these theories from being conclusive on the superconducting mechanism, which arises from a subtle energy balance between the Coulomb and the electron-phonon interactions in a multiorbital system. Therefore, in order to pin down the pairing mechanism, a non-empirical analysis is indispensable: If fully *ab initio* calculations quantitatively reproduced the phase diagram and $T_c$, they would provide precious microscopic information to identify unambiguously the superconducting mechanism, opening the path towards a predictive theory of superconductivity in correlated molecular materials.

In this work we carry out such non-empirical analysis using a combination of density functional theory (DFT) and dynamical mean-field theory (DMFT), an approach which accurately describes the properties of many correlated materials (22). DMFT is particularly accurate for compounds with short-ranged interactions and a large coordination number *z* such as $A_3C_{60}$, a highly symmetric three-dimensional system with *z*=12.

# RESULTS

**Theoretical phase diagram and comparison with experiments**

In Fig. 1 we show the bandstructure obtained within DFT for f.c.c. $Cs_3C_{60}$. Since the $t_{1u}$ bands crossing the Fermi level are well isolated from the other bands, we construct a lattice Hamiltonian consisting of the $t_{1u}$-electron and phonon degrees of freedom. The Hamiltonian is defined in the most general form,

$$\mathcal{H} = \sum_{ijk\sigma} \mathcal{H}_{ij}^{(0)}(\mathbf{k})\, c_{i\mathbf{k}}^{\sigma\dagger} c_{j\mathbf{k}}^{\sigma} + \sum_{\mathbf{qkk'}} \sum_{iji'j'} \sum_{\sigma\sigma'} U_{ij,i'j'}(\mathbf{q})\, c_{i\mathbf{k+q}}^{\sigma\dagger} c_{j'\mathbf{k'}}^{\sigma'\dagger} c_{i'\mathbf{k'+q}}^{\sigma'} c_{j\mathbf{k}}^{\sigma}$$
$$+ \sum_{ijk\sigma} \sum_{\mathbf{q}\nu} g_{ij}^{\nu}(\mathbf{k},\mathbf{q})\, c_{i\mathbf{k+q}}^{\sigma\dagger} c_{j\mathbf{k}}^{\sigma} (b_{\mathbf{q}\nu} + b_{-\mathbf{q}\nu}^{\dagger}) + \sum_{\mathbf{q}\nu} \omega_{\mathbf{q}\nu} b_{\mathbf{q}\nu}^{\dagger} b_{\mathbf{q}\nu}, \qquad (1)$$

on the f.c.c. lattice where each site represents a $C_{60}$ molecule. $\mathcal{H}^{(0)}, U, g$ and $\omega$ are the electron one-body Hamiltonian, the Coulomb interaction, the electron-phonon coupling, and the phonon frequency, respectively. They all have indices of the Wannier orbitals $i,j,i',j'$, spin $\sigma,\sigma'$, momentum $\mathbf{k},\mathbf{q}$, and the phonons also have a branch index $\nu$. We construct maximally localized Wannier orbitals (23) as the basis, in order to make the hopping and interaction parameters as short-ranged as possible. We determine *all* the above parameters by state-of-the-art *ab initio* techniques (Materials and Methods). In particular the recently developed constrained density-functional perturbation theory (cDFPT) (24) has enabled calculations of the phonon-related terms ($g_{ij}^{\nu}$ and $\omega_{\mathbf{q}\nu}$) where the effects of the high-energy bands are incorporated into the parameter values. To solve the *ab initio* model, we employ the extended DMFT (E-DMFT) (22), which combines the ability of DMFT to treat local interactions with the inclusion of non-local terms. To study the superconductivity, we introduce the anomalous Green's function, whose integral over frequency gives superconducting order parameter, into the E-DMFT equations.

Figure 2A shows the theoretical phase diagram as a function of temperature $T$ and volume per $C_{60}^{3-}$ anion ($V_{C_{60}^{3-}}$). As a function of $T$ and $V_{C_{60}^{3-}}$, we find a paramagnetic metal, a paramagnetic Mott insulator, and the *s*-wave superconducting phase where two electrons in the same orbital form the Cooper pairs. The three phases are characterized by a finite spectral weight at the Fermi level, a Mott gap, and non-zero anomalous Green's function and order parameter, respectively. The metal-insulator transition is of the first order below 80 K and a direct first-order superconductor-insulator transition takes place around $V_{C_{60}^{3-}} = 780$ Å$^3$ at low temperature. Between the blue-solid line $V_{c2}(T)$ and the black-dotted line $V_{c1}(T)$, we find both a metallic and an insulating solutions. The first-order transition line $V_c(T)$, where the free energies of the two solutions cross, is expected to be close to $V_{c2}(T)$ at low temperature as in the pure multiorbital Hubbard model (25).

In Fig. 2B we reproduce the experimental phase diagram (9) for f.c.c. $A_3C_{60}$ systems to highlight the impressive agreement with our calculations. In particular, the maximum $T_c$ of ~ 28 K in theory is comparable to that of experiment ~ 35 K (5,9). The critical volumes and the slope of the metal-insulator transition line are also consistent with experiment. It is remarkable that our fully *ab initio* calculation, without any empirical parameters, reproduces quantitatively the experimental phase diagram including the unconventional superconductivity and the Mott insulator.

**Unusual intramolecular interactions as a key for unconventional physics**

The success of our theory in reproducing the experimental phase diagram demonstrates the reliability of the whole scheme including the estimates of the interaction parameters. We are therefore in a position to disentangle the crucial ingredients leading to the proximity of *s*-wave superconductivity and a Mott insulator in the same diagram. The key quantities are the effective intramolecular interactions (intraorbital and interorbital density-density interactions $U_{\text{eff}} = U+U_V+U_{\text{ph}}$ and $U'_{\text{eff}} = U'+U'_V+U'_{\text{ph}}$, and exchange term $J_{\text{eff}} = J+J_{\text{ph}}$) between the $t_{1u}$ electrons, which are employed as input interaction parameters for the E-DMFT calculation. Here, $U$, $U'$ and $J$ are the intramolecular Coulomb interactions screened by the high-energy electrons while $U_V = U'_V$ and $J_V = 0$ ($U_{\text{ph}}$, $U'_{\text{ph}}$, and $J_{\text{ph}}$) represent dynamical screening contributions from the intermolecular Coulomb (electron-phonon) interactions. These dynamical screening effects make the effective interactions dependent on electron's frequency $\omega$.

The solid curves in Fig. 3 show $U_{\text{eff}}(\omega)$ and $U'_{\text{eff}}(\omega)$, where we find $U'_{\text{eff}}(\omega) > U_{\text{eff}}(\omega)$ up to $\omega \sim 0.2$ eV except a narrow region around $\omega = 0.1$ eV. This remarkable inversion of the low-energy interactions is associated with the negative value of $J_{\text{eff}}(0)$ discussed below, since the relation $U'_{\text{eff}}(\omega) \sim U_{\text{eff}}(\omega) - 2J_{\text{eff}}(\omega)$ holds. As is apparent from $U > U'$ and $U + U_V > U' + U'_V$ (plotted by dotted and dashed curves in Fig. 3), it is the phonon contribution ($U_{\text{ph}}$, $U'_{\text{ph}}$, and $J_{\text{ph}}$) that causes the inversion. In fact, $U_{\text{ph}}(\omega)$ and $U'_{\text{ph}}(\omega)$ have strongly $\omega$-dependent structures for $\omega \lesssim 0.2$ eV because of the intramolecular Jahn-Teller phonons with the frequencies up to ~ 0.2 eV (2), which are comparable to the $t_{1u}$ bandwidth ~ 0.5 eV.

Table I shows the static ($\omega=0$) values of the phonon-mediated interactions for various $A_3C_{60}$ compounds. For all the interactions, the phonon contribution has opposite sign with respect to $U$, $U'$, and $J$. While $|U_{ph}(0)|$ and $|U'_{ph}(0)|$ are much smaller than $U$ and $U'$ ~ 1 eV, $|J_{ph}(0)|$ ~ 51 meV is remarkably larger than that of the positive Hund's coupling $J$ ~ 34 meV (26). As a result, an effectively *negative* exchange interaction $J_{eff}(0) = J + J_{ph}(0)$ ~ −17 meV is realized. The unusual sign inversion of the exchange interaction (17,18,27,28) results from the molecular nature of the localized orbitals, which yields a small $J$, combined with a strong coupling between the Jahn-Teller phonons and $t_{1u}$ electrons (2), which enhances $|J_{ph}(0)|$.

**Microscopic mechanism of superconductivity**

We find, through a detailed analysis (Section F in Supplementary Materials), the unusual multiorbital interactions indeed drives the exotic *s*-wave superconductivity: (i) $U'_{eff} > U_{eff}$ and $J_{eff} < 0$ around $\omega = 0$ generate a singlet pair of electrons (leading to a Cooper pair) (19) which sit on the same orbital rather than on different orbitals as shown by $\langle n_{i\uparrow} n_{i\downarrow} \rangle > \langle n_{i\uparrow} n_{j\downarrow} \rangle$ in Fig. 4A. (ii) $J_{eff}$ further enhances the pairing through a coherent tunneling of pairs between orbitals [the Suhl-Kondo mechanism (29,30)].

As we have discussed above, these unusual multiorbital interactions are caused by phonons. In this sense, phonons are necessary to explain the superconductivity. However, strong electronic correlations also play a crucial role, marking a substantial difference from conventional phonon-driven superconductivity (17-19): The strong correlations heavily renormalize the electronic kinetic energy (from a bare value of ~ 0.5 eV) so that even the small difference (~ 33-37 meV) between $U'_{eff}$ and $U_{eff}$ is sufficient to bind the electrons into intraorbital pairs (19) [i.e., it helps the mechanism (i)]. As a result, $T_c$ increases with the correlation strength (see $T_c$ vs $V_{C_{60}^{3-}}$ in Fig. 2A). This analysis clearly demonstrates that the superconductivity in alkali-doped fullerides is the result of a synergy between electron-phonon interactions and electronic correlations (17,18).

**A peculiar low-spin Mott insulator**

Also the normal state displays remarkable consequences of the cooperative effect of the various interactions. Figure 4A shows the volume dependence of the intraorbital double occupancy $D$ and the size $S$ of the spin per molecule at $T$=40 K. Here we follow the metallic solution, expected to be stable in nearly whole coexistence region (25). In the Mott insulating region with the Mott gap in the spectral function (Fig. 4B), $S$ approaches 1/2 (Fig. 4A), which is of the low-spin state, in agreement with the experimental observation (4-9). Another feature characteristic to the present Mott transition is the increase of $D$ with the increase of the correlation strength, in sharp contrast with standard Mott transitions, in which $D$ is drastically reduced.

This phenomenology also descends from $U_{eff} < U'_{eff}$ and $J_{eff} < 0$, which prefer onsite low-spin configurations with two electrons on one orbital (Fig. 4D). In the metallic phase, the electron hopping tends instead to make all the different local configurations equally likely (Fig. 4C) (19). As the correlation increases, the electrons gradually lose their kinetic energy and hence the six low-spin configurations ($\{n_1,n_2,n_3\}$ = {2,1,0}, {0,2,1}, {1,0,2}, {2,0,1}, {1,2,0}, {0,1,2}) become majority (Fig. 4C), which leads to the decrease (increase) of $S$ ($D$). Eventually, in the Mott insulating phase, these configurations become predominant. Remarkably the six (210) configurations, in each of which the orbital degeneracy is lifted (Fig. 4D), remain indeed degenerate and no orbital ordering occurs. This unusual orbitally-degenerate Mott state is consistent with the experimental observation of a dynamical Jahn-Teller effect (10).

**Discussion**

We have carried out a fully *ab initio* study of the phase diagram of alkali-doped fullerides, in which all the interactions of a multi-orbital Hubbard model coupled with phonons are computed from first principles with the only information of the atomic positions. The model is then solved by means of E-DMFT. We identify that a negative Hund's coupling arising from the electron-phonon interaction and the related condition $U_{eff} < U'_{eff}$ which favors intraorbital pairs (17,18,27,28) underlie the existence of the superconducting state, its competition with the low-spin Mott state, and the whole phase diagram, which in turn remarkably reproduces the experimental observations.

This identifies a general condition under which a material can show a high-temperature electron-phonon superconducting phase which benefits from strong correlations (17), and suggests that a wider family of molecular conductors can exhibit a similar physics as long as an inverted Hund's exchange is stabilized by electron-phonon coupling. In view of the accuracy demonstrated for $A_3C_{60}$, we can apply the methodology of this manuscript to design other similar materials, possibly tailoring their properties to maximize the transition temperature.

## Materials and Methods

We apply the DFT+DMFT method (22) to f.c.c. $A_3C_{60}$. We start from the global band structure (Fig. 1) obtained by the DFT (see Section A in Supplementary Materials for the calculation conditions). Considering that the low-energy properties are governed by the isolated $t_{1u}$ bands, we derive from first principles the lattice Hamiltonian [Eq. (1)] for the $t_{1u}$ electrons and the phonons.

As the basis for the lattice Hamiltonian, we construct the maximally localized Wannier orbitals (23) $\{\phi_{i\mathbf{R}}\}$ from the $t_{1u}$ bands. We calculate the transfer integrals by $t_{ij}(\mathbf{R}) = \langle \phi_{i\mathbf{R}} | \mathcal{H}_{KS} | \phi_{j\mathbf{0}} \rangle$ with the Kohn-Sham Hamiltonian $\mathcal{H}_{KS}$. $\mathcal{H}_{ij}^{(0)}(\mathbf{k})$ in Eq. (1) is the Fourier transform of $t_{ij}(\mathbf{R})$, which well reproduces the original DFT band dispersion (Fig. 1). The Coulomb interaction in the low-energy model incorporates the screening effects from the high-energy bands, whereas the screening processes within the $t_{1u}$ bands are excluded to avoid a double counting since the latter processes are taken into account when we solve the model. Such partially-screened Coulomb parameters are calculated (26) with the constrained random phase approximation (cRPA) (31). Similarly, we calculate renormalized electron-phonon coupling $g_{ij}^{\nu}(\mathbf{k},\mathbf{q})$ and phonon frequency $\omega_{\mathbf{q}\nu}$ by applying the recently-developed cDFPT (24).

In the partition function for the lattice Hamiltonian [Eq. (1)] written in the coherent state path-integral formalism, we can integrate out the phonon degrees of freedom. This results in an electronic model with the additional electron-electron interaction ($U_{ph}$, $U'_{ph}$, and $J_{ph}$) mediated by phonons on top of the cRPA Coulomb interaction. The onsite phonon-mediated interaction $V_{ij,i'j'}$ (with $U_{ph} = V_{ii,ii}$, $U'_{ph} = V_{ii,jj}$, $J_{ph} = V_{ij,ij} = V_{ij,ji}$) is given, on the Matsubara axis, by (24)

$$V_{ij,i'j'}(i\Omega_n) = -\sum_{\mathbf{q}\nu} \tilde{g}_{ij}(\mathbf{q},\nu) \frac{2\omega_{\mathbf{q}\nu}}{\Omega_n^2 + \omega_{\mathbf{q}\nu}^2} \tilde{g}_{i'j'}^*(\mathbf{q},\nu),$$

where $\Omega_n = 2n\pi T$ is the bosonic Matsubara frequency and $\tilde{g}_{ij}(\mathbf{q},\nu) = \frac{1}{N_\mathbf{k}}\sum_\mathbf{k} g_{ij}^\nu(\mathbf{k},\mathbf{q})$ with the number $N_\mathbf{k}$ of **k**-points. Here, the sum over the phonon branches runs from 1 to 189. In practice, we omit the lowest 9 branches [$\nu$=1-9, which correspond to the acoustic modes, the librations, and the alkali-ion vibrations at the octahedral sites (20)]. This is justified since the coupling between these modes and the $t_{1u}$ electrons is tiny (2).

We solve the derived lattice Hamiltonian by means of the E-DMFT (22), which maps the Hamiltonian [Eq. (1)] onto a three-orbital single-impurity model embedded in noninteracting bath subject to self-consistent conditions. In the mapping, the E-DMFT incorporates the dynamical screening through the off-site interactions into the onsite interactions ($U_V$ and $U'_V$) in a self-consistent way. To solve the impurity model, we employ the continuous-time quantum Monte Carlo method based on the strong coupling expansion (32). Further detail can be found in Sections C and D in Supplementary Materials.

The frequency dependence of $U_V$ and $U'_V$ in Fig. 3 is calculated with the Padé analytic continuation. We perform the analytic continuation onto $\omega+i\eta$ with $\eta = 0.01$ eV from the data along the Matsubara axis at 40 K for $Cs_3C_{60}$ with $V_{C_{60}^{3-}} = 762$ Å$^3$. The spectral functions in Fig. 4B are calculated with the maximum entropy method.

## Supplementary Materials

Text (Sections A-F)

Fig. S1. Frequency dependence of partially screened Coulomb interactions for fcc $Cs_3C_{60}$ with $V_{C_{60}^{3-}} = 762$ Å$^3$

Fig. S2. Frequency dependence of the superconducting gap function at 10 K.

Table S1. Summary of input parameters for the E-DMFT calculations.

Table S2. Stability of superconducting (SC) solution at 10 K.

References(33-49)

**Acknowledgements:** We would like to thank K. Nakamura for valuable comments on the manuscript, and Y. Iwasa and Y. Kasahara for fruitful discussions and for providing us with the experimental phase diagram. We acknowledge useful discussions with P. Werner, T. Ayral, Y. Murakami, H. Shinaoka, N. Parragh, G. Sangiovanni, M. Imada, A. Oshiyama, A. Fujimori, P. Wzietek, H. Alloul, M. Fabrizio, E. Tosatti, and G. Giovannetti. Some of the calculations were performed at the Supercomputer Center, ISSP, University of Tokyo. **Funding:** Y.N. is supported by Grant-in-Aid for JSPS Fellows (No. 12J08652) from Japan Society for the Promotion of Science (JSPS), Japan. S. S. is supported by Grant-in-Aid for Scientific Research (No. 26800179), from JSPS, Japan. M. C. is supported by FP7/ERC through the Starting Grant SUPERBAD (Grant Agreement n. 240524) and by EU-Japan Project LEMSUPER (Grant Agreement n.283214). **Author contributions:** Y. N. performed the calculations. All the authors discussed the calculated results and wrote the manuscript. M. C. and R. A. conceived the project. **Competing interests:** The authors declare that there are no competing interests.

# Figures and Tables

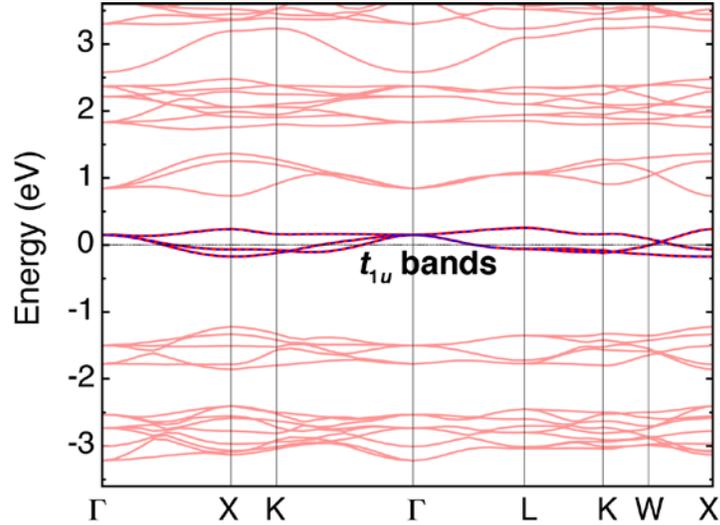

**Fig. 1. Band structure of f.c.c. Cs$_3$C$_{60}$ with $V_{C_{60}^{3-}} = 762$ Å$^3$.** The blue dotted curves represent the Wannier-interpolated band dispersion calculated from $\mathcal{H}_{ij}^{(0)}$ in Eq. (1).

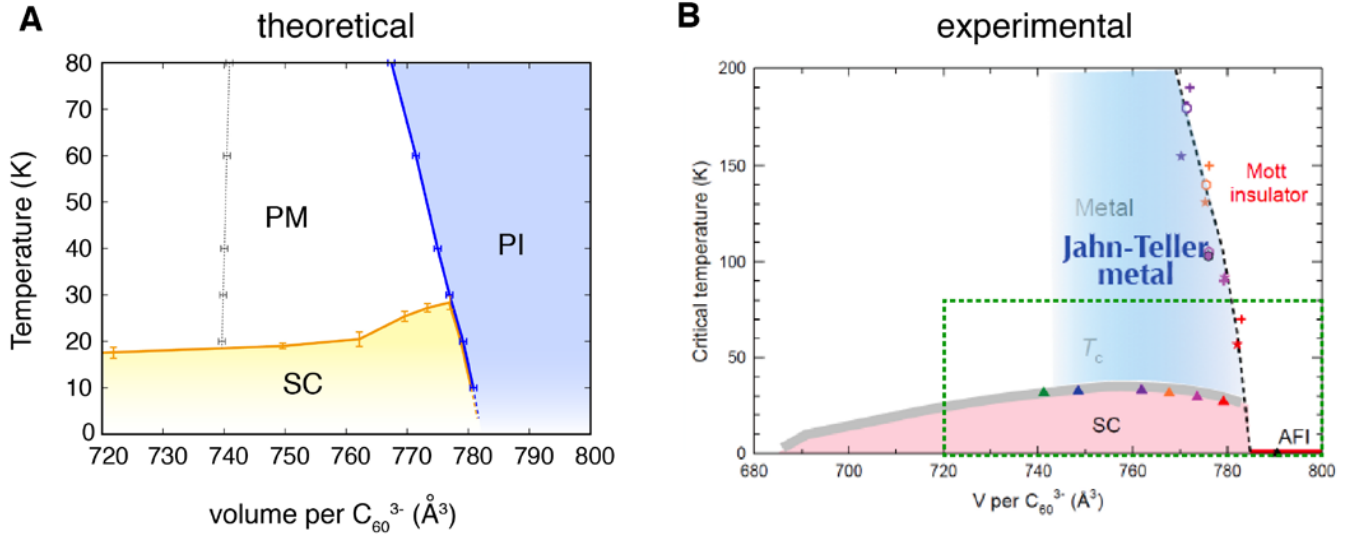

**Fig. 2. Theoretical and experimental phase diagrams.** (**A**) Phase diagram as a function of volume per $C_{60}$ molecule and temperature, obtained with the DFT+E-DMFT. PM, PI, and SC denote the paramagnetic metal, the paramagnetic insulator, and the superconducting phase, respectively. Metallic and insulating E-DMFT solutions coexist between the blue-solid line $V_{c2}(T)$ and the black-dotted line $V_{c1}(T)$. The error bars for $T_c$ originate from the statistical errors in the superconducting order parameters calculated with the quantum Monte Carlo method (see Section E in Supplementary Materials for details). The error bars for $V_{c1}(T)$ and $V_{c2}(T)$ are half the interval of volume grid in the calculation. For comparison, the experimental phase diagram [courtesy of Y. Kasahara (9)] is shown in the panel (**B**) where the region depicted in the panel (**A**) corresponds to the area surrounded by the green dotted lines. In the panel (**B**), AFI denotes the antiferromagnetic insulator.

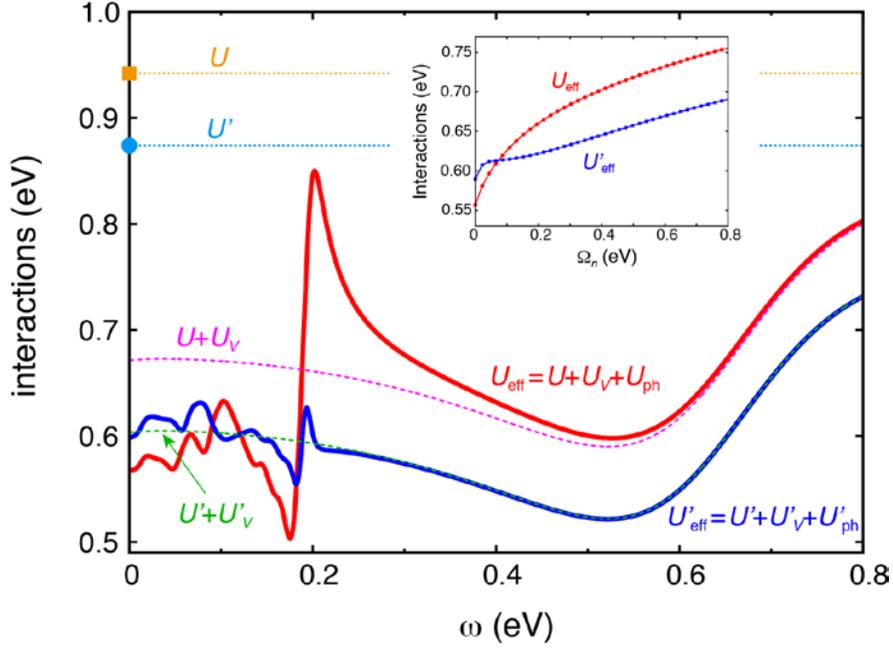

**Fig. 3. Frequency dependence of effective onsite interactions.** The effective intra and interorbital interactions ($U_{\text{eff}} = U + U_V + U_{\text{ph}}$ and $U'_{\text{eff}} = U' + U'_V + U'_{\text{ph}}$, respectively) consist of the cRPA onsite Coulomb repulsion ($U$, $U'$), the dynamical screening from the off-site interactions ($U_V = U'_V$), and the phonon-mediated interactions ($U_{\text{ph}}$, $U'_{\text{ph}}$). The data are calculated for $Cs_3C_{60}$ with $V_{C_{60}^{3-}} = 762\ \text{Å}^3$ at 40 K. We assume the cRPA Coulomb interactions to be static, whose validity is substantiated in Section B in Supplementary Materials. Inset: Frequency dependence of $U_{\text{eff}}$ and $U'_{\text{eff}}$ along the Matsubara frequency axis.

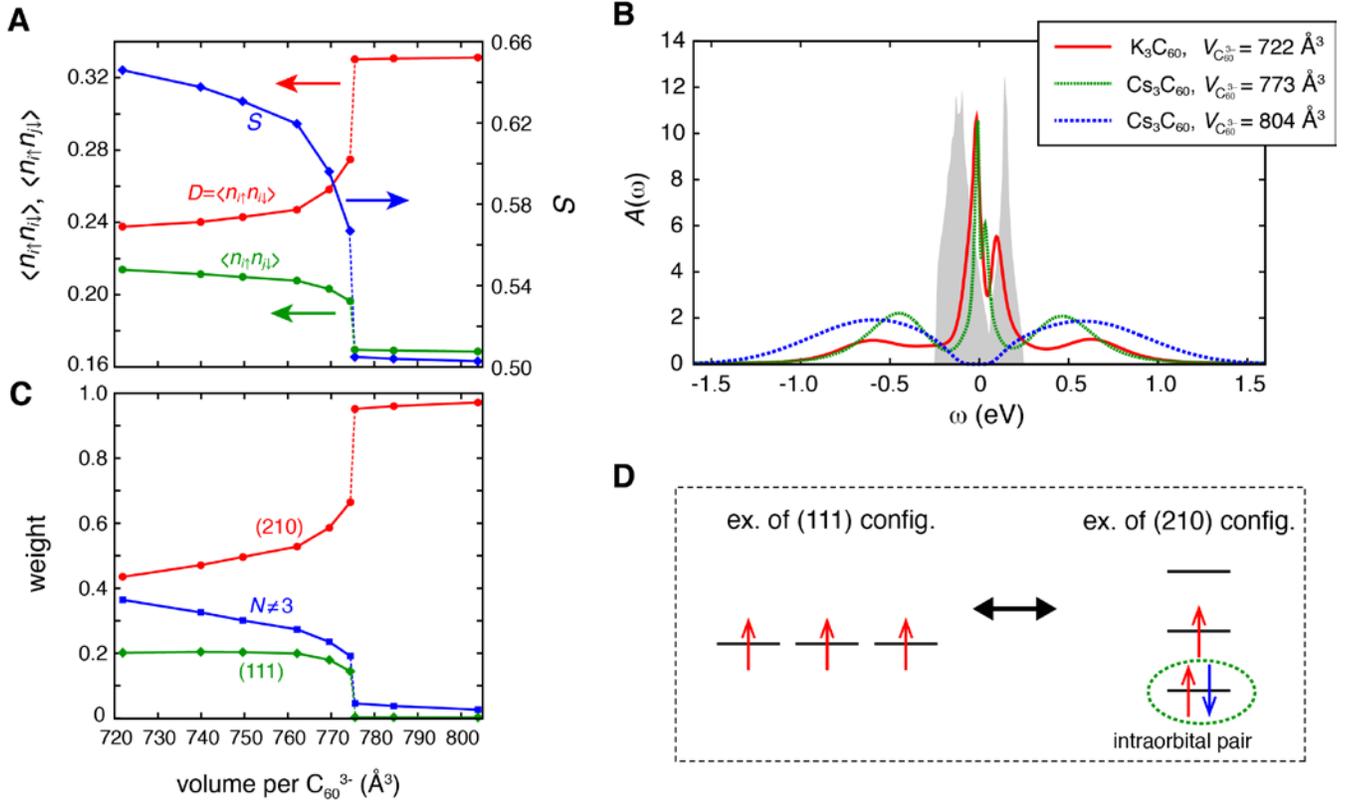

**Fig. 4. Double occupancy, size of spin, weights of intramolecular configurations, spectral functions at 40 K, and schematic pictures of representative intramolecular configurations.** (**A**) Volume dependence of the double occupancy $D = \langle n_{i\uparrow}n_{i\downarrow}\rangle$ (red), the interorbital interspin correlation $\langle n_{i\uparrow}n_{j\downarrow}\rangle$ (green), and the size $S$ of the spin per molecule (blue). (**B**) Spectral functions of several f.c.c. $A_3C_{60}$ systems at 40 K. For comparison, we show the DFT density of states for f.c.c. $K_3C_{60}$ ($V_{C_{60}^{3-}} = 722$ Å$^3$) as the shaded area. (**C**) Weights of several onsite configurations appearing in the quantum Monte Carlo simulations. (210) [(111)] generically denotes the configurations of $\{n_1,n_2,n_3\} = \{2,1,0\}, \{0,2,1\}, \{1,0,2\}, \{2,0,1\}, \{1,2,0\}, \{0,1,2\}$ [$\{n_1,n_2,n_3\} = \{1,1,1\}$] with $n_i$ being the occupation of orbital $i$. $N \neq 3$ with $N=n_1+n_2+n_3$ denotes the configurations away from half filling. (**D**) Illustrative pictures for the (111) and (210) configurations. The up and down arrows indicate the up- and down-spin electrons, respectively.

**Table 1. Material dependence of the static part of the phonon-mediated interactions.** $U_{ph}(0)$, $U'_{ph}(0)$, and $J_{ph}(0)$ are the phonon-mediated onsite intraorbital, interorbital, exchange interaction strengths between the $t_{1u}$ electrons at $\omega = 0$ calculated with the cDFPT. The energy unit is eV. The numbers just after the material names denote the volume occupied per $C_{60}^{3-}$ anion in Å$^3$.

|  | $U_{ph}(0)$ | $U'_{ph}(0)$ | $J_{ph}(0)$ |
|---|---|---|---|
| $K_3C_{60}$ (722) | −0.15 | −0.053 | −0.050 |
| $Rb_3C_{60}$ (750) | −0.14 | −0.042 | −0.051 |
| $Cs_3C_{60}$ (762) | −0.11 | −0.013 | −0.051 |
| $Cs_3C_{60}$ (784) | −0.12 | −0.022 | −0.051 |
| $Cs_3C_{60}$ (804) | −0.13 | −0.031 | −0.052 |

# Supplementary Materials

## A. Details of the model derivation part

We start from the DFT bandstructure calculations using QUANTUM ESPRESSO package (33). We adopt the local density approximation with the Perdew-Zunger parameterization (34) to approximate the exchange-correlation functional. Following the conditions used in Ref. 20, we prepare the Troullier-Martins norm-conserving pseudopotentials (35) in the Kleinman-Bylander representation (36) for the carbon and the alkali atoms. The configurations used to generate the pseudopotentials for C, K, Rb, and Cs are $(2s)^{2.0}(2p)^{2.0}$, $(3p)^{6.0}(4s)^{0.0}(3d)^{0.0}$, $(4p)^{6.0}(5s)^{0.0}(4d)^{0.0}$, and $(5p)^{6.0}(6s)^{0.0}(5d)^{0.0}$, respectively. The nonlinear core correction (37) is applied to the pseudopotentials for the alkali atoms. The relativistic effects are taken into account in the Rb and Cs pseudopotentials within the scalar relativistic approximation (38). The cutoff energy for the wave functions is set to be 50 Ry, and we employ 4×4×4 **k** points. With the above conditions, we perform the structure optimization with fixing the lattice constants to the values employed in Ref. 26 and ignoring the orientational disorder of $C_{60}$ molecules. Then, we calculate the band structure for the optimized structure.

We construct maximally localized Wannier orbitals (23) from the DFT $t_{1u}$ bands, and calculate the model parameters in the lattice Hamiltonian [Eq. (1)] as described in Methods. See Ref. 26 for further details of the calculations of the tight-binding parameters and the partially screened Coulomb interactions. The cDFPT calculations are performed with 2×2×2 **q**-mesh and a Gaussian smearing of 0.025 Ry. For the details of the cDFPT, see Ref. 24.

In Table S1, we summarize the values of the onsite interaction parameters for the five different fcc $A_3C_{60}$ systems. The effective interaction parameters employed in the subsequent E-DMFT calculation are given by the sum of the listed interactions (cRPA and phonon-mediated interactions) and the dynamical screening from off-site interactions. For comparison, we also show the $t_{1u}$ bandwidth $W$. In this low-energy model, $W$ determines the energy scale, the ratio $U_{\text{eff}}/W$ between the effective interaction ($U_{\text{eff}} = U+U_{\text{ph}}+U_V$) and the bandwidth ($W$) determines the strength of the effective mass renormalization, and $J_{\text{eff}}$ determines the strength of the pairing through mechanisms (i) and (ii) explained in the section "Microscopic mechanism of superconductivity" in the main text.

In drawing the theoretical phase diagram, we use a linear interpolation of the parameters for systems with $V_{C_{60}^{3-}} \neq$ 722, 750, 762, 784, 804 Å³ since the computation of the Coulomb and phonon parameters are quite demanding for the large unit cell, which contains 63 atoms. For example, the parameters for $V_{C_{60}^{3-}}$ = 736 Å³ are calculated as [(parameters for $K_3C_{60}$) + (parameters for $Rb_3C_{60}$)]/2. We expect that the errors in $T_c$ due to this linear interpolation are negligible since the $T_c$ curve in Fig. 2A is smooth against $V_{C_{60}^{3-}}$. For example, the change of $T_c$ between $K_3C_{60}$ with $V_{C_{60}^{3-}}$ = 722 Å³ and $Rb_3C_{60}$ with $V_{C_{60}^{3-}}$ = 750 Å³ is only ~ 1.5 K.

## B. Frequency dependence of partially screened Coulomb interactions

In this study, we assume that the partially screened Coulomb interactions are static, i.e., $U(\omega) = U(\omega=0)$, $U'(\omega) = U'(\omega=0)$, and $J(\omega) = J(\omega=0)$. This is because the calculation of the partially screened Coulomb interactions in the full frequency range for all the materials in Table S1 is highly expensive. In order to check the validity of the above assumption, we calculate the partially screened Coulomb interactions in a limited frequency range for one material, fcc $Cs_3C_{60}$ with $V_{C_{60}^{3-}}$ = 762 Å³. As we show below, the frequency dependence is indeed weak below a frequency much larger than the bandwidth.

We perform the cRPA calculation in the conditions described in Ref. 26. We expand the dielectric function in plane waves with the energy cutoff of 7.5 Ry. 335 (120 occupied+3 target+212 unoccupied) bands are considered in the calculation of the polarization function. The integral over the Brillouin-zone is evaluated with the generalized tetrahedron method (39,40). We employ a Lorentzian smearing for the delta function with the full width at half maximum of 0.3 eV.

Figure S1 shows the results of $U(\omega)$, $U'(\omega)$, $J(\omega)$. We find that they are almost flat in the frequency region where the phonon-mediated interactions $U_{ph}(\omega)$ and $J_{ph}(\omega)$ are active ($\omega \lesssim$ 0.2 eV). Therefore, the frequency dependences of the bare Coulomb interactions would not change the balance between the Coulomb and phonon-mediated interactions in Fig. 3. Furthermore, the inset of Fig. S1 shows that the values of the cRPA interactions differ by less than 15% from the $\omega$ = 0 value up to, at least, $\omega$ = 3 eV which is much larger than the bandwidth ~ 0.5 eV.

Finally, we discuss the origin of the small frequency dependence at low energy. In general, the values of the screened Coulomb interactions change drastically around the plasmon frequency. In $C_{60}$ compounds, the most prominent plasmon peak in the loss function is seen around $\omega = 27$ eV, which is the collective excitation involving all the valence electrons (41). This plasmon is largely responsible for the substantial reduction of the Coulomb interactions from their bare values ($U = 3.32$ eV, $U' = 3.12$ eV, and $J = 0.10$ eV) to those at $\omega = 0$ ($U = 0.94$ eV, $U' = 0.87$ eV, and $J = 0.035$ eV) (26). There also exists a less prominent plasmon peak at around $\omega = 6.5$ eV, which mainly involves the $\pi$ electrons (41). Since these plasmon frequencies (27 eV and 6.5 eV) are large, we see only a weak frequency dependence in Fig. S1.

## C. E-DMFT expression of dynamical screening through off-site Coulomb interactions

Within the E-DMFT (22,42-45), we take into account the effect of the off-site Coulomb interaction $\frac{1}{2}\sum_{lm} V_{lm}(N_l - \langle N_l \rangle)(N_m - \langle N_m \rangle)$, where $N_l$ denotes the operator of the total electron density at the $l$-th site. Here, we neglect the orbital dependence of the off-site Coulomb interaction because it is very weak (26). By means of the Hubbard-Stratonovich transformation, we recast the non-local interaction into a coupling between electrons and auxiliary bosons. Then, we map the resultant Hamiltonian onto a single-impurity Hamiltonian subject to the following self-consistent conditions (44,45):

$$\begin{cases} \mathbf{\Sigma}(i\omega_n) = \boldsymbol{\mathcal{G}}_0^{-1}(i\omega_n) - \boldsymbol{G}^{-1}(i\omega_n) \\ \boldsymbol{\mathcal{G}}_0^{-1}(i\omega_n) = \left\{ \frac{1}{N_\mathbf{k}} \sum_\mathbf{k} [(i\omega_n + \mu)\boldsymbol{I} - \boldsymbol{\mathcal{H}}_0(\mathbf{k}) - \mathbf{\Sigma}(i\omega_n)]^{-1} \right\}^{-1} + \mathbf{\Sigma}(i\omega_n) \end{cases} \quad (S1)$$

for the electron degrees of freedom and

$$\begin{cases} \Pi(i\Omega_n) = \mathcal{D}_0^{-1}(i\Omega_n) - D^{-1}(i\Omega_n) \\ \mathcal{D}_0^{-1}(i\Omega_n) = \left\{ \frac{1}{N_\mathbf{q}} \sum_\mathbf{q} \frac{1}{V_\mathbf{q}^{-1} - \Pi(i\Omega_n)} \right\}^{-1} + \Pi(i\Omega_n) \end{cases} \quad (S2)$$

for the auxiliary-boson degrees of freedom, respectively. Here, $\mathbf{\Sigma}$ ($\Pi$), $\boldsymbol{\mathcal{G}}_0$ ($\mathcal{D}_0$), $\boldsymbol{G}$ ($D$) denote the self-energy, the Weiss function, and the impurity-site Green's function for the electrons (auxiliary bosons), where the bold symbols denote matrices with respect to orbital indices. $\omega_n = (2n - 1)\pi T$ ($\Omega_n = 2n\pi T$) is the fermionic (bosonic) Matsubara frequency. $\mu$ is the chemical potential.

$\mathcal{H}_0(\mathbf{k})$ denotes the electron one-body Hamiltonian, which is the same as that in Eq. (1) in the main text. $V_\mathbf{q}$ is the Fourier transform of $V_{lm}$.

As in the case of the phonon-mediated interaction, we adopt the coherent-state path-integral representation for the partition function. Integrating out the auxiliary boson fields, we obtain the dynamical screening contribution $U_V(i\Omega_n) = U'_V(i\Omega_n) = \mathcal{D}_0(i\Omega_n)$ originating from the off-site interactions. As we see in Fig. 3 in the main text, $U_V$ and $U'_V$ reduce the value of the effective onsite interaction by ~ 0.27 eV. Even without this off-site screening contribution, we obtain qualitatively similar phase diagram as that depicted in Fig. 2A. This is because the essential conditions in realizing the remarkable phase diagram are the inversions of the usual inequalities of $J_\text{eff} > 0$ and $U_\text{eff} > U'_\text{eff}$, for which the off-site screening contribution does not play a role. However, the reduction helps to stabilize the metallic phase against the Mott insulating phase, which improves agreement between the theory and the experiment.

## D. Detail of the analysis of impurity problem

In order to solve the impurity problem in the E-DMFT, we employ the continuous-time quantum Monte Carlo method based on the strong coupling expansion (CT-HYB) (32). We deal with the dynamical density-density-type interactions $U_\text{eff}(i\Omega_n)$ and $U'_\text{eff}(i\Omega_n)$ in the scheme based on the Lang-Firsov transformation (46), which makes the computational cost for the dynamical interaction comparable to that for the static interaction (47). On the other hand, the non-density-type electron-phonon couplings and the non-density-type Coulomb interactions, such as the spin-flip and pair-hopping terms, have to be treated via the perturbation expansion (48). This is because the Lang-Firsov transformation becomes efficient only for the density-density-type interactions. In the alkali-doped fullerides, the negative exchange interaction ($J_\text{ph}$) mediated by phonons has a magnitude comparable to the positive exchange interaction ($J$). We find that, in such a situation, the perturbative expansion of $J$ and $J_\text{ph}$ leads to a severe negative sign problem. Since the alkali-doped fullerides have rather high phonon frequencies up to ~ 0.2 eV (2), we consider the phonons generating non-density-type couplings in the anti-adiabatic limit. The resultant instantaneous spin-flip and pair-hopping interactions $J_\text{eff} = J + J_\text{ph}(0)$ do not cause a serious negative-sign problem. The effect of dynamical spin-flip and pair-hopping interactions remains for future studies.

To summarize, the action $S_{\text{imp}}$ of the impurity problem, which we solve with the CT-HYB method, reads

$$S_{\text{imp}} = -\int_0^\beta d\tau\, d\tau' \sum_{ij\sigma} c_{i\sigma}^\dagger(\tau) [\mathcal{G}_0^{-1}(\tau-\tau')]_{ij} c_{j\sigma}(\tau') + \int_0^\beta d\tau \sum_i U \tilde{n}_{i\uparrow}(\tau)\tilde{n}_{i\downarrow}(\tau)$$

$$+ \int_0^\beta d\tau \sum_{i<j,\sigma} U' \tilde{n}_{i\sigma}(\tau)\tilde{n}_{j\bar\sigma}(\tau) + \int_0^\beta d\tau \sum_{i<j,\sigma} (U' - J_{\text{eff}})\tilde{n}_{i\sigma}(\tau)\tilde{n}_{j\sigma}(\tau)$$

$$+ \int_0^\beta d\tau \sum_{i\neq j} J_{\text{eff}}\, c_{i\uparrow}^\dagger(\tau) c_{j\downarrow}^\dagger(\tau) c_{i\downarrow}(\tau) c_{j\uparrow}(\tau) + \int_0^\beta d\tau \sum_{i\neq j} J_{\text{eff}}\, c_{i\uparrow}^\dagger(\tau) c_{i\downarrow}^\dagger(\tau) c_{j\downarrow}(\tau) c_{j\uparrow}(\tau)$$

$$+ \frac{1}{2}\int_0^\beta d\tau\, d\tau' \sum_i \tilde{n}_i(\tau)\big[U_{\text{ph}}(\tau-\tau') + U_V(\tau-\tau')\big]\tilde{n}_i(\tau')$$

$$+ \frac{1}{2}\int_0^\beta d\tau\, d\tau' \sum_{i\neq j} \tilde{n}_i(\tau)\big[U'_{\text{ph}}(\tau-\tau') + U'_V(\tau-\tau')\big]\tilde{n}_j(\tau') \qquad (S3)$$

where $\tilde{n}_{i\sigma} = c_{i\sigma}^\dagger c_{i\sigma} - \frac{1}{2}$ and $\tilde{n}_i = \tilde{n}_{i\uparrow} + \tilde{n}_{i\downarrow}$. The second, third, and fourth terms on the r.h.s. of Eq. (S3) denote the density-type interactions. The fifth and sixth terms correspond to the spin-flip and pair-hopping terms, respectively. The dynamical screening terms appear in the seventh and eighth terms.

### E. Method to determine $T_c$

The extended DMFT calculations are performed with the temperature mesh with the interval of 2.5 K. The lowest temperature reached in the calculation is 10 K. At each temperature, we calculate the dimensionless superconducting order parameter $P_{\text{SC}} = \sum_i \langle c_{i\downarrow} c_{i\uparrow}\rangle$. At 10 K, $P_{\text{SC}}$ takes a value $\sim 0.03$ in the superconducting region in Fig. 2A. To estimate $T_c$, we assume that the order parameter as a function of temperature behaves like $P_{\text{SC}}(T) \propto \sqrt{T_c - T}$ for $T \lesssim T_c$. We pick the highest and the second highest temperatures ($T_1$ and $T_2$, respectively), which give stable superconducting solutions ($P_{\text{SC}} \gtrsim 0.01$). Then, $T_c$ is determined by

$$T_c = T_1 + (T_1 - T_2)\frac{P_{\text{SC}}^2(T_1)}{P_{\text{SC}}^2(T_2) - P_{\text{SC}}^2(T_1)}. \qquad (S4)$$

The resulting $T_c$ agrees well with the temperature region where the convergence of the E-DMFT self-consistent loop is very slow, which is a sign of the vicinity to $T_c$. The error bars for $T_c$ in Fig. 2A originate from the statistical error in the values of $P_{\text{SC}}(T)$.

At 10 K, $P_{SC}$ shows a dome-like structure as a function of $V_{C_{60}^{3-}}$, which is consistent with the result for the three-band model (18). This dome-like structure is not seen in $T_c$ (Fig. 2A), probably because it is masked by the Mott insulating phase.

## F. Detailed analysis on superconducting mechanism

In order to identify crucial factors for the superconductivity, we perform several additional calculations with flexibly changing the input parameters for the E-DMFT from the *ab initio* values. The results are summarized in Table S2. First, we set the pair-hopping interaction to be zero keeping the other parameters unchanged, and restart the E-DMFT self-consistent loop from the converged superconducting solutions for several $V_{C_{60}^{3-}}$ at 10 K. The result is that the superconducting phase becomes unstable, implying that $T_c$ -if any- is below 10K in the absence of the pair-hopping term. This result clearly shows that the pair-hopping interaction is crucial for the superconductivity, despite its tiny value ($\sim -17$-$19$ meV) compared to e.g., the bandwidth $W \sim 0.5$ eV and Hubbard $U \sim 1$ eV. The pair-hopping interaction allows the Cooper pairs to move from one orbital to another. The enhancement of the superconductivity through the interband scattering is known as the Suhl-Kondo mechanism (29,30).

We also perform calculations with putting the spin-flip interaction to zero and keeping the other parameters unchanged. In this case, the superconducting solution survives, or is even slightly more stabilized, suggesting a minor role of the spin-flip terms.

Another important factor is the inequality $U_{\text{eff}}(i\Omega_n) < U'_{\text{eff}}(i\Omega_n)$ (49) around $\Omega_n = 0$, which is caused by the phonon-mediated interactions $U_{\text{ph}}$ and $U'_{\text{ph}}$. To examine the role of the inequality, we artificially set $U'_{\text{ph}}(i\Omega_n)$ to coincide with $U_{\text{ph}}(i\Omega_n)$ and restart the E-DMFT simulations from the stable superconducting solutions at 10 K with fixing the other parameters. In this setting, $U'_{\text{eff}}(i\Omega_n)$ is smaller than the original value, and the inequality $U_{\text{eff}}(i\Omega_n) > U'_{\text{eff}}(i\Omega_n)$ holds for all the Matsubara frequencies. The results at several $V_{C_{60}^{3-}}$ show no superconductivity, signaling that the present mechanism for superconductivity is radically different from the conventional scenario: In a standard BCS framework, the superconducting solution should survive or even be enhanced because the effective Coulomb interactions are reduced from the original values. Thus, the unusual inequality $U_{\text{eff}}(i\Omega_n) < U'_{\text{eff}}(i\Omega_n)$ is shown to be essential for the superconductivity. As is discussed in the main text, this condition favors, together with strong correlations, the formation of intraorbital pairs (19), which emerge as a "seed" of the strongly correlated superconductivity of alkali-doped fullerides.

Finally, we show the frequency dependence of the superconducting gap function $\Delta(i\omega_n)$ in Fig. S2. Here, we omit the orbital index since there is no orbital dependence in the gap function. We find that there exists a sign change in the gap function along the imaginary frequency axis. In the high frequency limit, the pairing interaction recovers a bare repulsive value $U + 2J_{\text{eff}}$. The repulsive interaction at high frequency favors the sign change in the gap function.

## Supplementary Figures and Tables

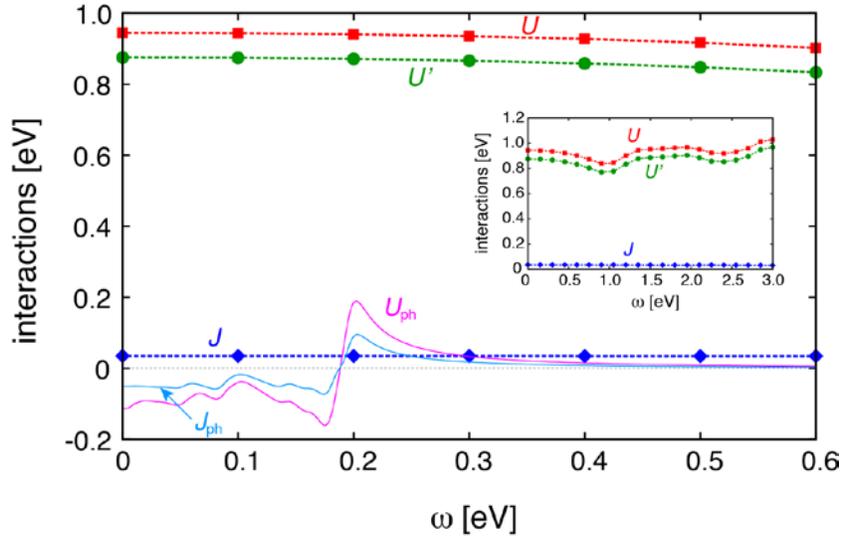

**Fig. S1. Frequency dependence of partially screened Coulomb interactions for fcc $Cs_3C_{60}$ with $V_{C_{60}^{3-}} = 762$ Å$^3$.** $U(\omega)$, $U'(\omega)$, and $J(\omega)$ are the intraorbital, interorbital, and exchange components of the cRPA Coulomb interactions, respectively. The frequency dependences in a wider frequency region are shown in the inset. The weak structure around $\omega = 1.0$ (2.5) eV originates from the transition processes between $t_{1u}$ bands and $t_{1g}$ ($h_u$) bands, respectively. For comparison, we also plot the frequency dependence of the phonon-mediated intraorbital and exchange interactions [$U_{\text{ph}}(\omega)$ and $J_{\text{ph}}(\omega)$, respectively].

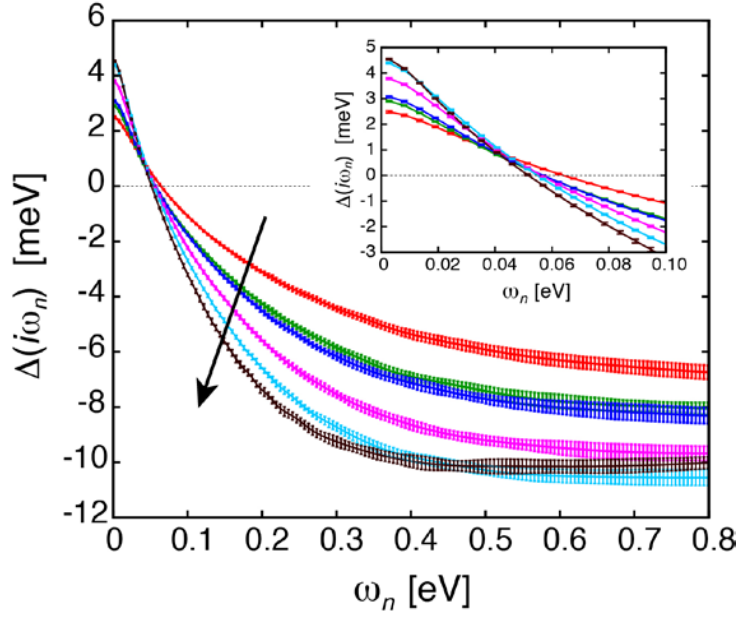

**Fig. S2. Frequency dependence of the superconducting gap function at 10 K.** We plot the gap function $\Delta(i\omega_n)$ for six materials with different lattice constants. Along the direction of the arrow in the figure, the lattice constant increases: The corresponding $V_{C_{60}^{3-}}$'s are $V_{C_{60}^{3-}}$ = 722 (red), 750 (green), 762 (blue), 767 (purple), 773 (light blue), and 779 (dark brown) Å$^3$. The data in the low frequency region are zoomed in the inset.

**Table S1. Summary of input parameters for the E-DMFT calculations.** $U$ and $U'$ ($U_{ph}$ and $U'_{ph}$) are the onsite intra and interorbital Coulomb (phonon-mediated) interactions, respectively. $J_{eff} = J + J_{ph}(0)$ is the effective exchange and pair-hopping interactions with $J$ and $J_{ph}(0)$ being the Hund's coupling and the static part of the phonon-mediated exchange interaction, respectively. For comparison, we list the values of the DFT bandwidth $W$ of the $t_{1u}$ bands, too.

|  | $V_{C_{60}^{3-}}$ (Å$^3$) | $W$ (eV) | $U$ (eV) | $U'$ (eV) | $U_{ph}(0)$ (meV) | $U'_{ph}(0)$ (meV) | $J_{eff}=J+J_{ph}(0)$ (meV) |
|---|---|---|---|---|---|---|---|
| K$_3$C$_{60}$ | 722 | 0.502 | 0.826 | 0.764 | −152 | −53 | −18.5 |
| Rb$_3$C$_{60}$ | 750 | 0.454 | 0.915 | 0.847 | −143 | −42 | −16.5 |
| Cs$_3$C$_{60}$ | 762 | 0.427 | 0.942 | 0.874 | −114 | −13 | −16.5 |
| Cs$_3$C$_{60}$ | 784 | 0.379 | 1.014 | 0.945 | −124 | −22 | −16.5 |
| Cs$_3$C$_{60}$ | 804 | 0.341 | 1.068 | 0.997 | −134 | −31 | −16.5 |

**Table S2. Stability of superconducting (SC) solution at 10 K.** We show the property of the E-DMFT solution in each case when we change the interaction parameters from the *ab initio* values.

| *ab initio* | (pair hopping)=0 | (spin flip)=0 | $U'_{ph}(i\Omega_n) = U_{ph}(i\Omega_n)$ |
|---|---|---|---|
| SC | Non SC | SC | Non SC |